\documentclass[12pt]{a_t}
\usepackage{cmap}
\usepackage{graphicx}
\usepackage{multicol}
\usepackage[justification=centering, labelsep=period]{caption}
\usepackage{float}
\usepackage{xhfill}

\begin{document}  

\year{2020}
\title{МЕТОД ПРИТЯГИВАЮЩИХ ЦИЛИНДРОВ. РЕШЕНИЕ ОБЩЕЙ ЛИНЕЙНОЙ ЗАДАЧИ СЛЕЖЕНИЯ}

\thanks{Результаты разделов 3 и 4 получены при поддержке гранта Российского научного фонда (№ 18-79-10104) в ИПМаш РАН. Результаты разделов 5 и 6 получены при поддержке гранта Президента Российской Федерации (грант № МД-1054.2020.8).}

\authors{А.А. ПЕРЕГУДИН (peregudin@itmo.ru) \\ (Национальный исследовательский университет ИТМО, \\ Институт проблем машиноведения РАН, Санкт-Петербург)}

\maketitle

\begin{abstract}
Представлен метод притягивающих цилиндров -- обобщение метода инвариантных эллипсоидов {\color{black} на случаи задач слежения и наблюдения}. На основе разработанного метода предложен алгоритм расчета параметров регулятора, {\color{black} обеспечивающего ограниченность ошибки слежения или наблюдения в присутствии ограниченных внешних возмущений. Эффективность} предложенного подхода продемонстрирована на примерах.
\end{abstract}

\it Ключевые слова: \rm задача слежения, задача наблюдения, подавление возмущений.
\section{Введение}

Задача подавления ограниченных внешних возмущений на основе метода инвариантных эллипсоидов 
ранее рассматривалась в [1-3]. В частности, в \cite{inv1} решалась задача стабилизации возмущенной системы с измеряемым состоянием с помощью 
статического регулятора; в \cite{inv2} аналогичный подход применялся для стабилизации возмущенной системы с измеряемым выходом -- 
к статическому регулятору был добавлен динамический наблюдатель Люенбергера; в \cite{inv3} аналогичная задача была решена с помощью динамического регулятора общего вида. Однако во всех этих работах решалась только задача стабилизации объекта, но не задача слежения.

Распространению метода инвариантных эллипсоидов на задачу слежения посвящены работы \cite{track1,track2}, в которых на систему наложен ряд дополнительных условий. Так, в \cite{track1} предполагается, что все компоненты 
задающего воздействия измеряемы и могут быть использованы 
регулятором, а в \cite{track2} дополнительно предполагается, что их производные также измеряемы и ограничены.
Отметим, что в обеих работах состояние объекта является измеряемым, а используемый регулятор -- статическим. В этом смысле работы \cite{track1,track2} обобщают \cite{inv1}, но не более поздние \cite{inv2,inv3}, в которых уже используется динамический регулятор при неизмеряемом состоянии объекта.

{\color{black} Целью настоящей работы является обобщение метода инвариантных эллипсоидов на случай задачи слежения при использовании динамического регулятора по выходу.} В качестве инструмента используется новый метод, основанный на притягивающих множествах более общего вида, в том числе неограниченных по части переменных.

В работе использованы следующие обозначения: $\mathbb{S}^n = \{ A \in \mathbb{R}^{n \times n} \; | \; A^{\rm T} = A\}$,  если $A \in \mathbb{R}^{m \times n}$, то $\operatorname{range}A = \{ Ax \in \mathbb{R}^m \; | \; x \in \mathbb{R}^n\}$, $\operatorname{ker}A = \{ x \in \mathbb{R}^n \; | \; Ax = 0\}$, \linebreak $A^+$ -- псевдообратная Мура-Пенроуза (существует у любой матрицы вне зависимости от полноты ранга).

\section{Мотивирующий пример}

\color{black}В качестве мотивирующего примера рассмотрим неустойчивую систему, состояющую из объекта первого порядка и наблюдателя первого порядка, заданную как
\begin{equation} \label{motivation}
\left\{
\begin{aligned}
&\dot{x}(t) = ax(t)+bf(t), \\
&\dot{\widehat{x}}(t) = (a-l)\widehat{x}(t)+lx(t), 
\end{aligned}
\right.
\end{equation} \color{black}
где $x(t)\in \mathbb{R}$ -- состояние объекта, $\widehat{x}(t)\in \mathbb{R}$ -- состояние наблюдателя, $f(t)\in \mathbb{R}$ -- внешнее возмущение, $|f(t)|\le 1$, $b>0$ и $l>a>0$. {\color{black}В качестве выходной переменной объединенной системы \eqref{motivation} рассмотрим ошибку наблюдения}
\begin{equation*}
    y(t) = x(t) - \widehat{x}(t).
\end{equation*} \color{black}
Представленная система неустойчива и не имеет притягивающего эллипсоида по состоянию. Однако в силу того, что выход $y(t)$ имеет устойчивую динамику
\begin{equation*}
\dot{y}(t) = (a-l)y(t)+bf(t),
\end{equation*}
у системы существуют притягивающие эллипсоиды по выходу. Минимальный притягивающий эллипсоид по выходу (в данном случае -- отрезок минимальной длины) может быть найден аналитически как
\begin{equation} \label{otrezok}
\Big\{ y \in \mathbb{R} \; \Big| \; y^2 \le \frac{b^2}{(a-l)^2}\Big\},
\end{equation}
однако в пространстве состояний ему соответствует \it неограниченное \rm множество
\begin{equation} \label{strip}
\Big\{(x,\widehat{x}) \in \mathbb{R}^2 \; \Big| \;|x-\widehat{x}| \le \frac{b}{l-a}\Big\},
\end{equation}
не являющееся эллипсоидом. Метод инвариантных эллипсоидов \cite{inv1} не может быть непосредственно применен для поиска притягивающего множества \eqref{otrezok}, потому как у системы \eqref{motivation} не существует инвариантного эллипсоида по состоянию. Существует, однако, притягивающее подмножество пространства состояний системы в форме полосы \eqref{strip}. Траектория попадает в эту полосу и затем движется в ней, неограниченно удаляясь от начала координат. Рисунок 1 иллюстрирует описанную ситуацию. 

\begin{figure}[H]
\centering
\includegraphics[width=0.42\textwidth]{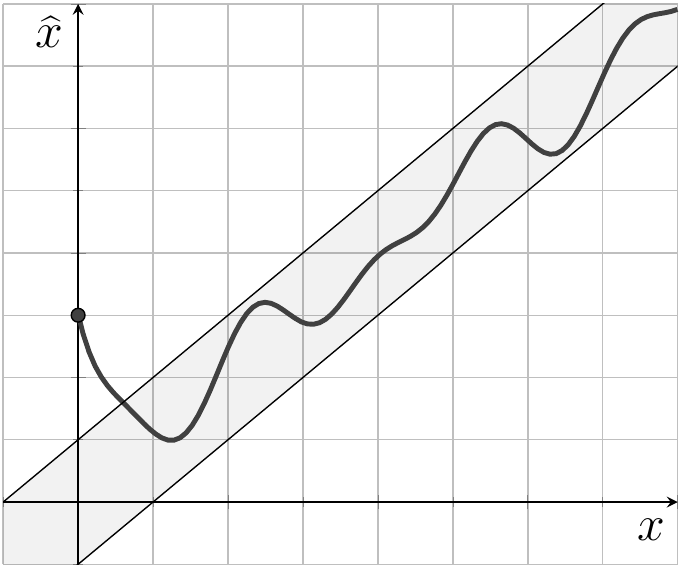}
\caption{Пример траектории системы. Серым цветом выделено неограниченное притягивающее подмножество пространства состояний.}
\end{figure}

{\color{black}Данный} пример показывает, что для решения задачи слежения или  наблюдения в общем случае недостаточно метода инвариантных (притягивающих) эллипсоидов, разработанного в [1-3] для решения задач стабилизации. {\color{black}Для обобщения существующего метода на случаи, подобные представленному выше, в настоящей работе развивается \it метод притягивающих цилиндров. \rm}

\section{Геометрические основы метода притягивающих цилиндров}

Для описания рассматриваемых в настоящей работе притягивающих подмножеств введем следующее определение.

\begin{definition}  Подмножество пространства $\mathbb{R}^n$, заданное как 
\begin{equation}
\label{opr1}
\Big\{x \in \mathbb{R}^n \; \Big| \; x^{\rm T}Qx \le 1\Big\},
\end{equation} где $Q \in \mathbb{S}^n$, $Q \succeq 0$ и $\operatorname{rank} Q = k$, называется $(k,n)$-цилиндром. 
\end{definition}

Примеры $(k,n)$-цилиндров приведены на рис. 2-4. Им соответствуют множества $\Big\{(x,y) \in \mathbb{R}^2 \; \Big| \; x^2 \le 1\Big\}$, $\Big\{(x,y,z) \in \mathbb{R}^3 \; \Big| \; x^2+(y-z)^2 \le 1\Big\}$, $\Big\{(x,y,z) \in \mathbb{R}^3 \; \Big| \; z^2 \le 1\Big\}$, каждое из которых можно задать в виде \eqref{opr1}, выбрав соответствующую матрицу $Q$.  
\begin{figure}[H]
\minipage{0.24\textwidth}
  \centering
  \includegraphics[scale = 0.7]{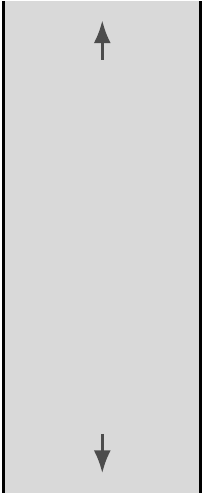}
  \caption{Бесконечная полоса -- пример $(1,2)$-цилиндра.}\label{fig:awesome_image1}
\endminipage\hfill
\minipage{0.32\textwidth}
  \centering
  \includegraphics[scale = 0.7, angle = -35]{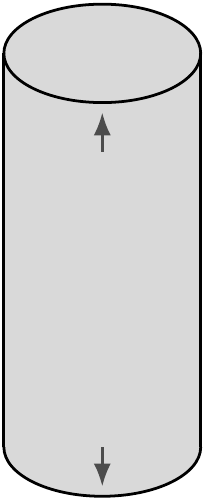}
  \caption{Бесконечный цилиндр -- пример $(2,3)$-цилиндра.}\label{fig:awesome_image2}
\endminipage\hfill
\minipage{0.32\textwidth}%
  \centering
  \vspace{1cm}
  \includegraphics[scale = 0.7]{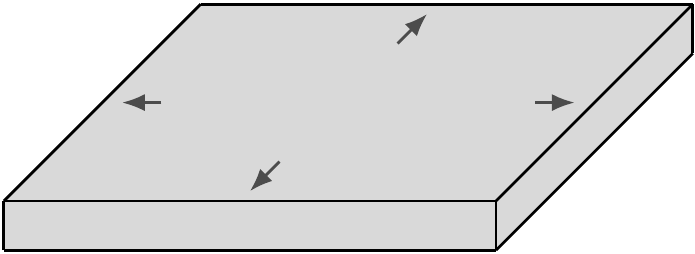}
  \vspace{0.9cm}
  \caption{Слой пространства между двумя плоскостями -- пример $(1,3)$-цилиндра.}\label{fig:awesome_image3}
\endminipage
\end{figure}

\begin{remark}
Отметим, что при $k=n$ множество \eqref{opr1} является эллипсоидом и что в [1-3] авторами рассматривались притягивающие подмножества именно такого вида. При $k<n$ множество \eqref{opr1} эллипсоидом уже не является, но при этом также может являться притягивающим подмножеством пространства состояний некоторой системы. В этом смысле метод притягивающих цилиндров является обобщением метода инвариантных (притягивающих) эллипсоидов.
\end{remark}

Сформулируем геометрическое описание $(k,n)$-цилиндров.

\begin{statement} \label{st1}
Как множество в линейном пространстве $(k,n)\text{-цилиндр}$ \eqref{opr1} является суммой $k$-мерного эллипсоида, лежащего в подпространстве $\operatorname{range}Q$, и всего подпространства $\operatorname{ker}Q$. 
\end{statement}

\begin{corollary} \label{cor1}
Как топологическое пространство  $(k,n)$-цилиндр гомеоморфен прямому произведению замкнутого $k$-мерного шара и $\mathbb{R}^{n-k}$.
\end{corollary}

Доказательства утверждения 1 и следствия 1 приведены в Приложении.

Известно, что образом эллипсоида при линейном отображении является эллипсоид. Обобщим это утверждение на случай $(k,n)$-цилиндров.

\begin{statement} \label{pq}
Пусть $C \in \mathbb{R}^{m \times n}$, $\operatorname{rank} C = m$. Образом $(k,n)$-цилиндра \eqref{opr1} при отображении $y=Cx$ является $(r,m)$-цилиндр 
\begin{equation*}
    \Big\{y \in \mathbb{R}^m \; \Big| \; y^{\rm T}Ry \le 1\Big\}, \quad R = C^{+ \rm T}M \big( I - (MN) (MN)^+ \big)MC^{+},
\end{equation*}
где $r = \operatorname{rank} R$, $M=Q^{1/2}$, $N = I-C^+C$.
\end{statement}

\begin{corollary} \label{pq_plane}
При $n \ge 2$ проекцией $(k,n)$-цилиндра на произвольную плоскость является либо вся плоскость, либо полоса (часть плоскости между двумя параллельными прямыми), либо эллипс (часть плоскости, ограниченная эллипсом).
\end{corollary}

Доказательства утверждения 2 и следствия 2 приведены в Приложении.

Изложенные выше утверждения дают читателю базовые представления о геометрии рассматриваемых в настоящей работе притягивающих подмножеств.

\section{Анализ. Метод притягивающих цилиндров}

Рассмотрим линейную динамическую систему
\begin{equation}
\label{anal1}
\dot{x}(t) = Ax(t) + Bf(t),
\end{equation}
где $x(t) \in \mathbb{R}^n$ -- вектор состояния, $f(t) \in \mathbb{R}^m$ -- внешнее возмущение, $A$, $B$ -- вещественные матрицы соответствующих размерностей. Предполагается, что внешнее возмущение ограничено и что известна матрица $G \succ 0$ такая, что 
\begin{equation} \label{anal2}
    f^{\rm T}(t)Gf(t) \le 1, \quad \forall t \ge 0.
\end{equation}

\begin{definition} 
Притягивающим $(k,n)$-цилиндром системы \eqref{anal1}-\eqref{anal2} называется множество \eqref{opr1} такое, что для всех траекторий системы выполнено
\begin{equation*} 
    \limsup_{t \, \rightarrow \, +\infty} x^{\rm T}(t)Qx(t) \le 1,
\end{equation*}
а также, если $x(t_0) \in \eqref{opr1}$, то  $x(t) \in \eqref{opr1}$ при всех $t \ge t_0$.
\end{definition}

Согласно определению 2, притягивающие $(k,n)$-цилиндры являются одновременно притягивающими и инвариантными подмножествами пространства состояний. 

Сформулируем достаточное условие существования притягивающего цилиндра в пространстве состояний системы \eqref{anal1}-\eqref{anal2}.

\begin{theorem} \label{thm:1}
Если матрица $C \in \mathbb{R}^{k \times n}$ такова, что 
\begin{equation*} 
\operatorname{rank} C = k, \quad CA(I-C^+C)=0, 
\end{equation*}
и если существуют  $P \succ 0$ и $\alpha>0$ такие, что
\begin{equation*}
\begin{bmatrix}
PCAC^{+}+(CAC^{+})^{\rm T}P+\alpha P & PCB \\ (CB)^{\rm T}P & -\alpha G
\end{bmatrix} \prec 0,
\end{equation*}
то подмножество $\{x \in \mathbb{R}^n  \; | \;  x^{\rm T}C^{\rm T}PCx \le 1\}$ пространства состояний системы \eqref{anal1}-\eqref{anal2} является притягивающим $(k,n)$-цилиндром. 
\end{theorem}

Доказательство теоремы 1 приведено в Приложении.

\begin{remark}
Отметим, что в частном случае, при $C = I \in \mathbb{R}^{n \times n}$, теорема 1 \linebreak совпадает с основным результатом работы [1], и тогда притягивающий $(k,n)$-цилиндр является инвариантным эллипсоидом.
\end{remark}

\pagebreak

\section{Синтез. Общая линейная задача слежения}

\subsection{\color{black} Постановка задачи}

{\color{black}Рассмотрим объект управления}
\begin{equation}
\label{plant}
\left\{
\begin{aligned}
\dot{x}(t)&=A_1x(t)+B_1u(t)+C_1w(t), \\
y(t) &= D_1x(t)+ E_1u(t)+ F_1w(t), \\
\end{aligned}
\right.
\end{equation}
где $x(t) \in \mathbb{R}^{a_1}$ -- неизмеряемый вектор состояния, $u(t) \in \mathbb{R}^{b_1}$ -- управление, $w(t) \in \mathbb{R}^{c_1}$ -- {\color{black}неизмеряемый вектор возмущений}, $y(t) \in \mathbb{R}^{b_2}$ -- измеряемый выход, $A_1$,$B_1$,$C_1$,$D_1$,$E_1$,$F_1$ -- известные вещественные матрицы соответствующих размерностей.

{\color{black}Пусть эталонная модель задана как}
\begin{equation}
\label{reference}
\left\{
\begin{aligned}
\dot{x}_r(t)&=A_2x_r(t)+C_2h(t), \\
g(t) &= D_2x_r(t), \\
\end{aligned}
\right.
\end{equation}
где $x_r(t) \in \mathbb{R}^{a_2}$ -- неизмеряемый вектор состояния, $h(t) \in \mathbb{R}^{c_2}$ -- неизмеряемое {\color{black} задающее} воздействие, $g(t) \in \mathbb{R}^{c_1}$ -- измеряемый {\color{black} эталонный} выход,  $A_2,C_2,D_2$ -- известные вещественные матрицы соответствующих размерностей.

\begin{remark}
Формально, эталонную модель можно не выделять в отдельную систему, а сделать частью уравнения \eqref{plant}, объединив вектор $x(t)$ с вектором $x_r(t)$, $w(t)$ с $h(t)$ и $y(t)$ с $g(t)$. {\color{black} Однако в настоящей работе предлагается рассмотреть \eqref{plant} и \eqref{reference} как две различные модели. Такое разделение не является обязательным, но помогает лучше понять смысл целевой переменной $z$, раскрываемый далее.}
\end{remark}

Для достижения цели управления предполагается использование регулятора заданного динамического порядка $a_3 \in \mathbb{N} \cup \{0\}$ вида 
\begin{equation}
\label{controller}
\left\{
\begin{aligned}
\dot{x_c}(t)&=A_3x_c(t)+B_3y(t)+C_3g(t), \\
u(t) &= D_3x_c(t) + E_3y(t) + F_3g(t), \\
\end{aligned}
\right. 
\end{equation}
где $x_c(t) \in \mathbb{R}^{a_3}$ -- вектор состояния регулятора, $A_3,B_3,C_3,D_3,E_3,F_3$ -- вещественные матрицы соответствующих размерностей, подлежащие выбору. 

\begin{remark} 
При $a_3 = 0$ матрицы $A_3,B_3,C_3,D_3$ пустые, регулятор \eqref{controller} является статическим и описывается формулой $u(t) = E_3y(t) + F_3g(t)$. Поскольку современные компьютерные программы (например, \textsc{matlab}) свободно работают с пустыми матрицами различных размерностей, далее не будем рассматривать этот случай отдельно, считая его частью общей теории.
\end{remark}

Предполагается, что внешние сигналы $w(t)$, $h(t)$ ограничены и что известна матрица $G \succ 0$ такая, что 
\begin{equation}
\label{bounds}
\begin{bmatrix}
w(t) \\ h(t)
\end{bmatrix}^{\rm T} G \begin{bmatrix}
w(t) \\ h(t)
\end{bmatrix} \le 1, \quad \forall t \ge 0.
\end{equation}

\begin{remark} {\color{black}
Консервативность основного результата, {\color{black}который будет изложен в разделе 5.2,} может быть несколько снижена путем замены \eqref{bounds} на пару ограничений $w(t)^{\rm T}G_1 w(t) \le 1$, $h(t)^{\rm T}G_2h(t) \le 1$, где $G_1, G_2 \succ 0$, или даже на большее число ограничений, наложенных на отдельные компоненты данных векторов. Однако это приведет к увеличению числа свободных переменных и  усложнению формулировок, не обязательному для настоящей статьи.}
\end{remark}

Цель управления формулируется следующим образом: при $w(t), h(t) \equiv 0$ обеспечить асимптотическую сходимость целевой переменной
\begin{equation}
\label{goal}
z(t) = K_1 x(t) + K_2 x_r (t) + K_3  x_c (t)
\end{equation}
к нулю. Если же $w(t), h(t) \not \equiv 0$, но выполнено условие \eqref{bounds}, то переменная \eqref{goal} должна асимптотически сходиться к ограниченному множеству вида $\{ z \in \mathbb{R}^k \; | \; z^{\rm T}Pz \le 1 \}$ и должна быть найдена соответствующая матрица $P \succ 0$. Предполагается, что матрицы $K_1 \in \mathbb{R}^{k \times a_1}$, $K_2 \in \mathbb{R}^{k \times a_2}$, $K_3 \in \mathbb{R}^{k \times a_3}$, задающие цель управления, известны и что $\operatorname{rank} \begin{bmatrix} K_1 & K_2 & K_3\end{bmatrix} = k$. Последнее условие наложено для удобства и не является ограничительным, так как его выполнения всегда можно добиться, убрав из матрицы $\begin{bmatrix} K_1 & K_2 & K_3\end{bmatrix}$ линейно зависимые строки.

\color{black} Сформулированная таким образом задача в рамках настоящей статьи называется \it общей линейной задачей слежения\rm.  Отметим несколько частных случаев: 

\begin{enumlist}
\item[1.]
Задача стабилизации. \rm Если $K_1 = I$, $K_2 = K_3= 0$, то при отсутствии внешних воздействий цель управления принимает вид $ \Vert x(t) \Vert \rightarrow 0$. В \cite{inv3} показано, что такая задача может быть решена с помощью метода инвариантных эллипсоидов.
\item[2.]
Задача слежения. \rm Если $K_1 = I$, $K_2 = -I$, $K_3 = 0$, то при отсутствии внешних воздействий цель управления имеет вид $\Vert x(t)-x_r(t) \Vert \rightarrow 0$, что соответствует слежению вектора состояния объекта за вектором состояния эталонной модели.
\item[3.]
Задача наблюдения. \rm Если $K_1 = I$, $K_2 = 0$, $K_3 = -I$, то регулятор \eqref{controller} превращается в наблюдатель, вектор состояния которого должен сходиться к вектору состояния объекта. Если внешние воздействия отсутствуют, то такая цель управления может быть сформулирована как $\Vert x_c(t)-x(t) \Vert \rightarrow 0$.
\end{enumlist}

Если матрицы $K_{i}$ выбраны иным образом, то цель управления представляет собой некоторое сочетание задач стабилизации, слежения и наблюдения (возможно, по части переменных). Таким образом, поставленная задача синтеза может быть интерпретирована как одна из этих трех базовых задач либо как их комбинация.

Отметим, что на матрицы $A_1, B_1, C_1, D_1, F_1, A_2, C_2, D_2$  в \eqref{plant} и \eqref{reference} не накладывается никаких ограничений. Однако требуется наложить ограничение на матрицу $E_1$, связанное с корректностью рассматриваемой обратной связи. Рассмотрим вспомогательный измеряемый выход $\widehat{y}(t) = \color{black} y(t) - E_1u(t) = \color{black} D_1x(t)+  F_1w(t)$   системы \eqref{plant}, который получается из выражения для $y(t)$, если убрать слагаемое с $E_1$.  Предположим, что поставленная задача может быть решена с помощью регулятора \eqref{controller}, в котором вместо выхода $y(t)$ используется вспомогательный выход $\widehat{y}(t)$. Тогда после подстановки $\widehat{y}(t) = y(t) - E_1u(t)$ получим закон управления в виде
\begin{equation*} 
u(t) = (I+E_1 E_3)^{-1} (D_3 x_c(t) + E_3y(t) + F_3g(t)) = \widehat{D}_3 x_c(t) + \widehat{E}_3y(t) + \widehat{F}_3g(t), 
\end{equation*}
для реализации которого необходимо, чтобы матрица $(I+E_1 E_3)^{-1}$ существовала. Именно это условие {\color{black} и накладывается} дополнительно. {\color{black} С учетом указанного свойства при формулировке основного результата достаточно ограничиться случаем $E_1 = 0$, что и будет сделано, при этом соответствующий регулятор для общего случая всегда может быть восстановлен. }

\subsection{Основной результат}

Введем вспомогательные обозначения для описания замкнутой системы
\begin{equation*}
A = 
\begin{bmatrix}
A_1 & 0 & 0 \\ 0 & A_2 & 0 \\ 0 & 0 & 0
\end{bmatrix}, \quad
B = 
\begin{bmatrix}
0 & B_1 \\ 0 & 0 \\ I & 0 
\end{bmatrix}, \quad
C = 
\begin{bmatrix}
C_1 & 0  \\ 0 & C_2  \\ 0 & 0 
\end{bmatrix}, \quad 
D = 
\begin{bmatrix}
0 & 0 & I \\ D_1 & 0 & 0 \\ 0 & D_2 & 0
\end{bmatrix}, 
\end{equation*}
\vspace{0cm}
\begin{equation*}
F = 
\begin{bmatrix}
0 & 0 \\ F_1 & 0 \\ 0 & 0 
\end{bmatrix}, \quad
X = 
\begin{bmatrix}
A_3 & B_3 & C_3 \\ D_3 & E_3 & F_3
\end{bmatrix}, \quad s(t) = \begin{bmatrix}
    x(t) \\ x_r(t) \\ x_c(t)
    \end{bmatrix}, \quad f(t) = \begin{bmatrix}
    w(t) \\ h(t)
    \end{bmatrix}, 
\end{equation*}
\vspace{0cm}
\begin{equation*}
M = A+BXD, \quad N = C+BXF,
 \quad 
   K = \begin{bmatrix}
    K_1 & K_2 & K_3
    \end{bmatrix}, \quad n = a_1+a_2+a_3.
\end{equation*}
Тогда уравнение замкнутой системы \eqref{plant}-\eqref{controller}, \eqref{goal} может быть записано как
\begin{equation}\label{closed}
\left\{
\begin{aligned}
    \dot{s}(t) &= Ms(t) + Nf(t), \\
    z(t) &= Ks(t),
\end{aligned}
\right.
\end{equation}
при этом ограничение \eqref{bounds} примет вид 
\begin{equation}\label{closed-bounds}
    f^{\rm T}(t)Gf(t) \le 1,\quad \forall t \ge 0.
\end{equation}

Перед формулировкой основного результата введем обозначения
\begin{equation} \label{notations}
\begin{gathered}
    H_1 = KAK^+ + KB(KB)^+KA(D(K^+K-I))^+DK^+,\\
    H_2 = KC + KB(KB)^+KA(D(K^+K-I))^+F,\\
    H_3 = KB, \quad H_4 = DK^+ +D(D(K^+K-I))^+DK^+,\\
    H_5 = F + D(D(K^+K-I))^+F.
\end{gathered}
\end{equation}
{\color{black} Сформулируем основной результат в виде теоремы.}
\begin{theorem} \label{thm:2}

Если матрицы $A,B,D,K$ таковы, что 
\begin{equation} \label{x_condition}
    KB(KB)^+KA(D(I-K^+K))^+D(I-K^+K) = KA(I-K^+K), 
\end{equation}
и если существуют  $P,Q \succ 0$,  $\mu_1, \mu_2 \in \mathbb{R}$, $\alpha > 0$ такие, что $PQ = I$ и
\begin{equation} \label{PQlmi}
\begin{aligned}
    \begin{bmatrix}
    H_1 Q + Q H_1^{\rm T} + \alpha Q & H_2 \\ H_2^{\rm T} & - \alpha G
    \end{bmatrix} &\prec \mu_1 \begin{bmatrix}
    H_3 H_3^{\rm T} & 0 \\ 0 & 0 
    \end{bmatrix}, 
    \\ \vspace{-1.6cm} 
    \begin{bmatrix}
    P H_1  + H_1^{\rm T} P + \alpha P & P H_2 \\ H_2^{\rm T} P  & - \alpha G
    \end{bmatrix} &\prec \mu_2 \begin{bmatrix}
    H_4^{\rm T} H_4 & H_4^{\rm T} H_5 \\ H_5^{\rm T} H_4 & H_5^{\rm T} H_5
    \end{bmatrix},
\end{aligned}
\end{equation}
то существует набор $X$ параметров регулятора \eqref{controller} такой, что подмножество $\{s \in \mathbb{R}^{n} \; | \;  s^{\rm T}K^{\rm T}PKs \le 1\}$ пространства состояний замкнутой системы \eqref{closed} является притягивающим $(k,n)$-цилиндром. 

При фиксированных $\alpha$, $P$ соответствующая матрица $X$ находится как
\begin{equation} \label{x_solution}
    X = (KB)^+KA(D(K^+K-I))^++Y+(KB)^+KBYD(D(K^+K-I))^+,
\end{equation}
где {\color{black} $Y$ --  любое решение линейного матричного неравенства}
\begin{equation} \label{Ylmi}
    \begin{bmatrix}
    P H_1  + H_1^{\rm T} P + \alpha P & P H_2 \\ H_2^{\rm T} P  & - \alpha G
    \end{bmatrix} + \begin{bmatrix}
    PH_3 Y H_4 + (PH_3 Y H_4)^{\rm T} & PH_3 Y H_5 \\ (PH_3 Y H_5)^{\rm T} & 0 
    \end{bmatrix}
    \prec 0.
\end{equation}
\end{theorem}

Доказательство теоремы 2 приведено в Приложении. Из доказательства, в частности,  следует, что при выполнении условий теоремы матричное неравенство \eqref{Ylmi} всегда имеет решение. 

Условие \eqref{x_condition} является вполне естественным: оно показывает, как должны соотноситься между собой параметры объекта \eqref{plant} и эталонной модели \eqref{reference}, чтобы решение соответствующей задачи слежения было возможным. Можно показать, что для задач стабилизации и наблюдения, при $K = \begin{bmatrix} I & 0 & 0 \end{bmatrix}$ и $K = \begin{bmatrix} I & 0 & -I \end{bmatrix}$ соответственно, условие \eqref{x_condition} всегда выполнено независимо от параметров объекта и эталонной модели. Если же $K = \begin{bmatrix} I & -I & 0 \end{bmatrix}$, что соответствует задаче слежения, то  условие \eqref{x_condition} представляется как 
\begin{equation*}
    B_1 B_1^{+} (A_1-A_2) (D_1^{\rm T}D_1+D_2^{\rm T}D_2)^{+}(D_1^{\rm T}D_1+D_2^{\rm T}D_2)=(A_1-A_2).
\end{equation*}
Следует отметить, что выполнение этого условия еще не означает возможность построения соответствующего регулятора, ведь помимо него должно быть также выполнено второе условие  теоремы 2, связанное с неравенствами \eqref{PQlmi}.

\subsection{Вычислительные аспекты}

Поскольку на матрицы $P$ и $Q$, входящие в формулировку теоремы 2, наложено дополнительное ограничение $PQ=I$, матричные неравенства \eqref{PQlmi} не являются линейными даже при фиксированном $\alpha$ и не могут быть непосредственно решены с помощью стандартных программных средств, таких как \textsc{yalmip} или \textsc{cvx}. {\color{black}Однако существует} эффективный алгоритм решения матричных неравенств именно с таким типом нелинейности, {\color{black} который в литературе можно встретить под названием <<Cone complementarity algorithm>> (алгоритм восполнения конуса) [6-9]}.

\begin{algorithm} \label{alg:1}

\ 

\begin{enumlist}[.] 

\item
Задать $i=0$. Зафиксировать параметр $\alpha > 0$ и найти положительно определенное решение $(\widehat{P}, \widehat{Q})$ системы линейных матричных неравенств \eqref{PQlmi}.  

\item
Положить $(R_i, S_i) = (\widehat{P}, \widehat{Q})$. 

\item
Найти положительно определенное решение $(\widehat{P}, \widehat{Q})$ задачи минимизации: 
\begin{equation*}
\begin{aligned}
    &\text{минимизировать }  &&\operatorname{trace} (PS_i + QR_i) \\ &\text{при условиях } && \begin{bmatrix}
    P & I \\ I & Q
    \end{bmatrix} \succeq 0, \; \text{\eqref{PQlmi}}.
    \end{aligned}
\end{equation*}
\item
Если выполнено условие остановки (например, величина $\operatorname{trace} (\widehat{P}S_i + \widehat{Q}R_i)$ достаточно близка к $2k$ или величина $i$ достигла предельного значения), перейти к шагу 5. Иначе увеличить $i$ на единицу и перейти к шагу 2. 

\item 
Принять полученную пару $(\widehat{P},\widehat{Q})$ в качестве пары $(P,Q)$, приближенно удовлетворяющей условиям теоремы 2. Закончить.
\end{enumlist}
\end{algorithm}

В \cite{alg1} показано, что алгоритм 1 сходится к паре $(P,Q)$, удовлетворяющей условиям \eqref{PQlmi} и доставляющей локальный минимум величине $\operatorname{trace} (PQ+QP)$. Несмотря на то, что в общем случае выполнение условия $PQ=I$ не гарантировано, алгоритм широко применяется [6-9] для решения задач теории управления, в которых возникают подобные матричные неравенства.

Применение теоремы 2 на практике во всей полноте предполагает выполнение следующей последовательности действий:

\begin{enumlist}[.]

\item 
Проверить, что для поставленной задачи выполнено условие \eqref{x_condition}. 

\item
Применив алгоритм 1, найти взаимообратные $P$ и $Q$, удовлетворяющие \eqref{PQlmi}.

\item 
Используя найденную матрицу $Q$, найти матрицу $Y$, удовлетворяющую \eqref{Ylmi}.

\item 
Вычислить матрицу $X$ параметров регулятора по формуле \eqref{x_solution}.

\end{enumlist}

Однако следует отметить, что нахождение пары взаимообратных матриц $(P,Q)$ не является обязательным для синтеза регулятора.  Если в процессе выполнения алгоритма 1 найдена матрица $P$ такая, что  неравенство \eqref{Ylmi} имеет решение $Y$, то нет необходимости далее выполнять алгоритм -- можно сразу перейти к вычислению параметров регулятора по формуле \eqref{x_solution}.

\section{Примеры}
\subsection{Задача слежения}

Рассмотрим объект управления \eqref{plant} с матрицами
\begin{equation*}
   A_1 = \begin{bmatrix}
            -2{,}99 & 3{,}10 \\
            -2{,}10 & 2{,}01
         \end{bmatrix},
    \quad 
    B_1 = \begin{bmatrix}
            1{,}5 \\ 1
            \end{bmatrix},
    \quad 
    C_1 =  \begin{bmatrix}
            0{,}15 \\ 0{,}15 
            \end{bmatrix},
\end{equation*}
\begin{equation*}
   D_1 = \begin{bmatrix}
            1 & -1
         \end{bmatrix},
    \quad 
    E_1 = \begin{bmatrix}
            0
            \end{bmatrix},
    \quad 
    F_1 =  \begin{bmatrix}
           0
           \end{bmatrix}
\end{equation*}
и эталонную модель \eqref{reference} с матрицами
\begin{equation*}
   A_2 = \begin{bmatrix}
            0{,}01 & 0{,}1 \\
            -0{,}1 & 0{,}01
         \end{bmatrix},
    \quad 
    C_2 = \begin{bmatrix}
            0
           \end{bmatrix},
    \quad 
    D_2 =  \begin{bmatrix}
            1 & -1 
            \end{bmatrix}.
\end{equation*}
Заметим, что объект управления является устойчивым, а эталонная модель, напротив, неустойчива. Условие \eqref{bounds} предполагается выполненным при $G = 1$.

Пусть динамический порядок $a_3$ регулятора \eqref{controller} положен равным 1, а целевая переменная \eqref{goal} задана как
\begin{equation*}
    z(t) = x(t) - x_r(t),
\end{equation*}
что соответствует задаче слежения и выбору матрицы
\begin{equation*}
    K = \begin{bmatrix}
    1 & 0 & -1 & 0 & 0 \\
    0 & 1 & 0 & -1 & 0
    \end{bmatrix}.
\end{equation*}
Нетрудно проверить, что в этом случае условие \eqref{x_condition} выполнено. Тогда в результате выполнения алгоритма 1 при $\alpha = 0{,}5$ может быть найдена матрица
\begin{equation*}
    P = \begin{bmatrix}
    \phantom{-}1485 &  -1585\\
    -1585 & \phantom{-}1698
    \end{bmatrix} \succ 0
\end{equation*}
и затем по формулам \eqref{Ylmi}, \eqref{x_solution} восстановлены параметры 
\begin{equation*}
    X = \begin{bmatrix}
A_3 & B_3 & C_3 \\ D_3 & E_3 & F_3
\end{bmatrix} = \begin{bmatrix}
0 & 0 & 0 \\
0 & -2{,}95 & 4{,}95
\end{bmatrix}
\end{equation*}
регулятора \eqref{controller}, который, несмотря на заданный динамический порядок $a_3=1$, оказывается статическим.

При моделировании внешнее возмущение было задано как $w(t) = \sin(0{,}4t)$, а начальные условия положены равными $x(0) = \begin{bmatrix} 0 & 0 \end{bmatrix}^{\rm T}$ и $x_r(0) = \begin{bmatrix} 1 & 1\end{bmatrix}^{\rm T}$.

Результаты моделирования представлены на рис. 5-6. На рис. 5 показана проекция траектории замкнутой системы: по оси абсцисс отложено значение первой координаты вектора состояния объекта, а по оси ординат -- значение первой координаты вектора состояния эталонной модели. Также показаны границы полосы, являющейся проекцией притягивающего $(k,n)$-цилиндра на эту плоскость.

\begin{figure}
\minipage{0.5\textwidth}
  \centering
  \includegraphics[width = \textwidth]{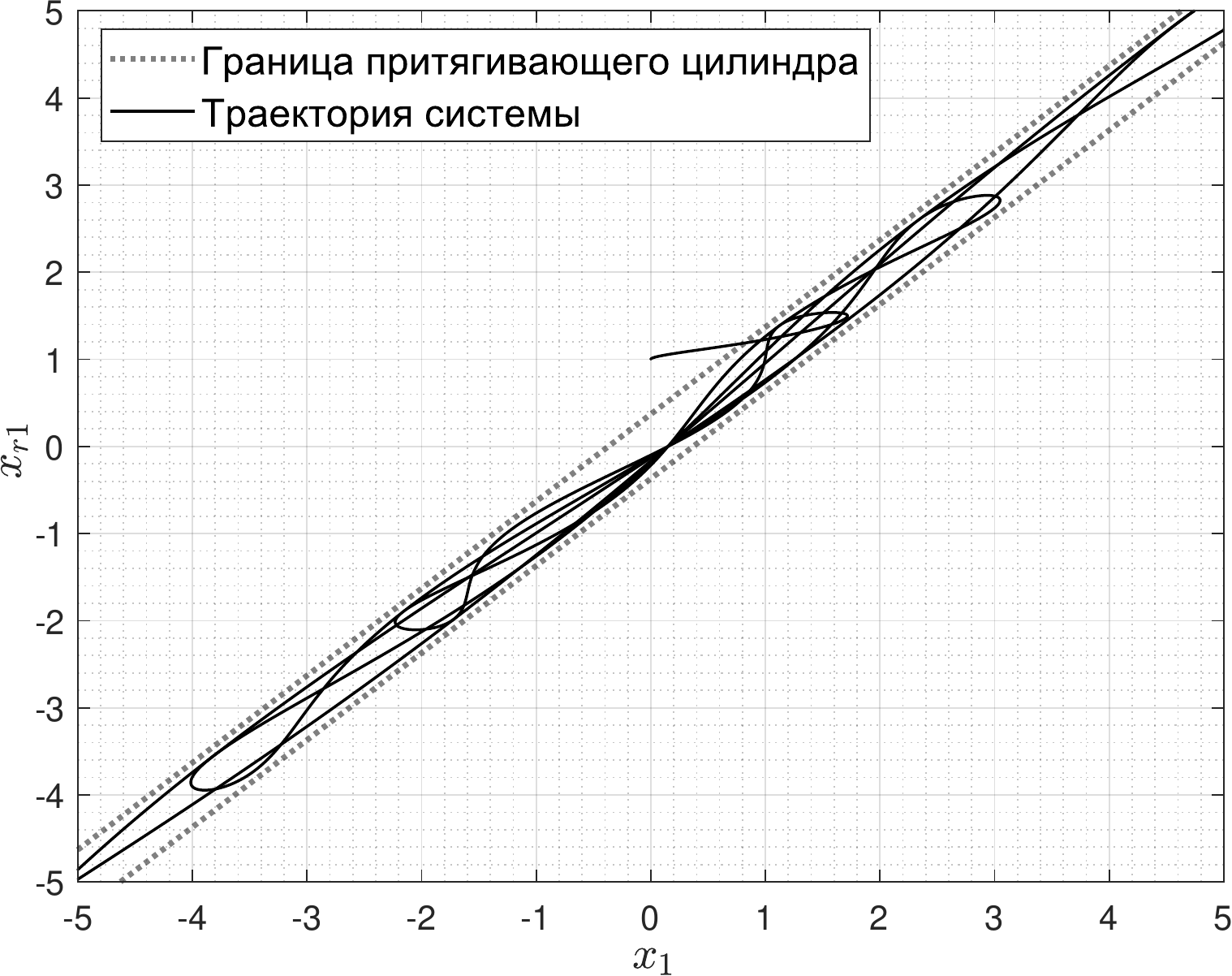}
  \caption{Проекция притягивающего $(k,n)$-цилиндра и траектории замкнутой системы.}
\endminipage\hfill
\minipage{0.5\textwidth}
  \centering
  \includegraphics[width = 0.93\textwidth]{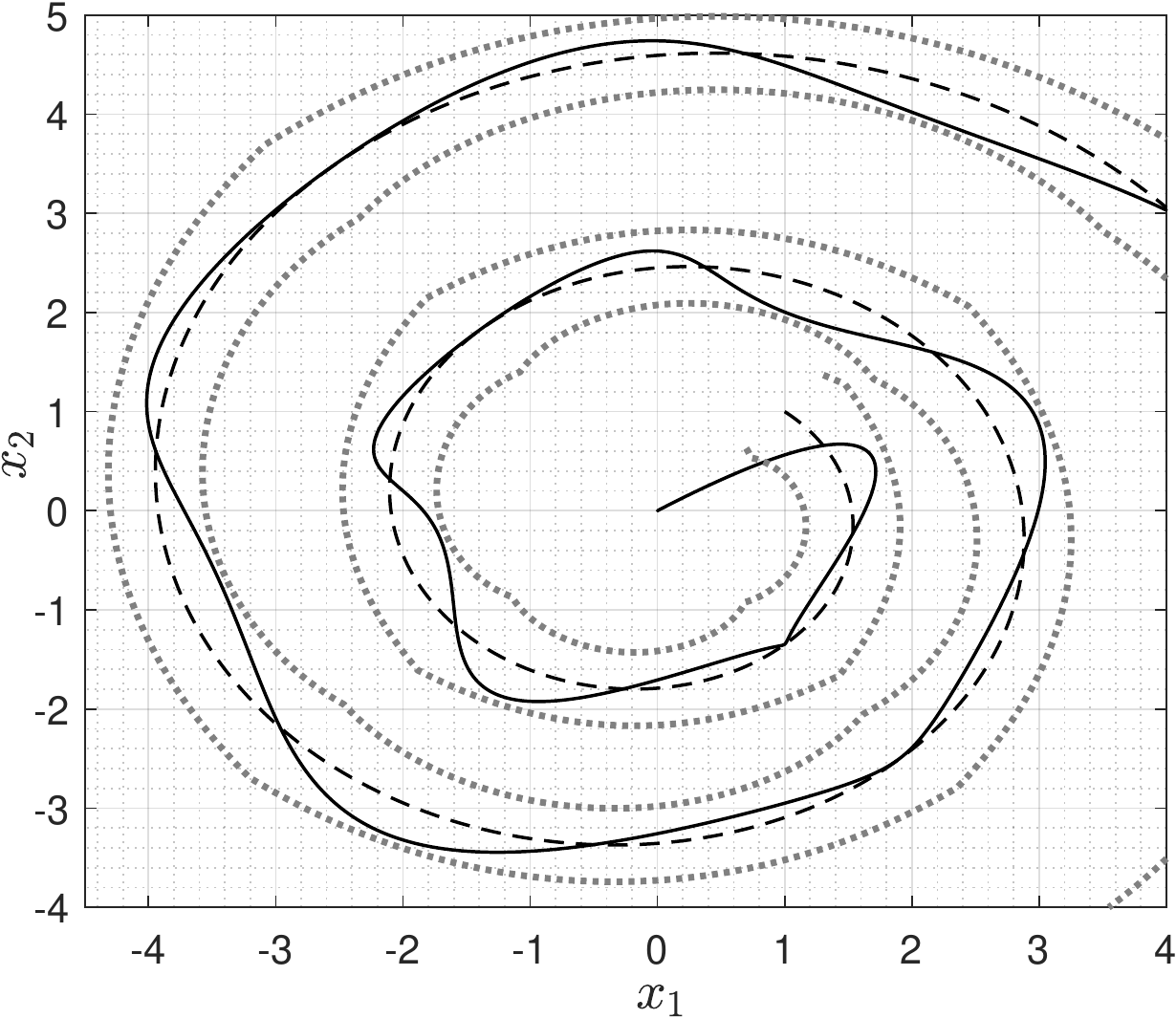}
  \caption{Траектория объекта (\rule[.5ex]{2em}{.4pt}), траектория эталонной модели (\rule[.5ex]{.5em}{.4pt} \rule[.5ex]{.5em}{.4pt} \rule[.5ex]{.5em}{.4pt}) и 
  границы допустимого коридора (\makebox[2em]{\protect\xdotfill{.4pt}}\!).}
\endminipage
\end{figure}

На рис. 6 показаны траектория $x(t)$ объекта и траектория $x_r(t)$ эталонной модели. На рисунке также отмечены границы <<допустимого коридора>>, имеющего следующий смысл: если в данный момент времени каждая из проекций траектории замкнутой системы -- на плоскость $(x_1,x_{r1})$ (см. рис. 5) и на плоскость $(x_2,x_{r2})$ (не приводится, так как выглядит аналогично) -- находится внутри соответствующей проекции притягивающего $(k,n)$-цилиндра, то траектория $x(t)$ в данный момент времени находится в границах допустимого коридора. Иными словами, тот факт, что траектория объекта начиная с какого-то момента не выходит за границы точечного пунктира, является иллюстрацией того, что траектория замкнутой системы попадает внутрь $(k,n)$-цилиндра и не покидает его. 

\subsection{Задача наблюдения}

Рассмотрим объект \eqref{plant} с матрицами
\begin{equation*}
   A_1 = \begin{bmatrix}
        0{,}168 &  -0{,}132 &  -0{,}052 \\
        0{,}148 &  -0{,}152 &   \phantom{-}0{,}028 \\
        0{,}204 &  -0{,}196 &  -0{,}006 \\
         \end{bmatrix},
    \quad 
    B_1 = \begin{bmatrix}
            0 \\ 0 \\ 0
            \end{bmatrix},
    \quad 
    C_1 =  \begin{bmatrix}
            0 \\ 0 \\ 0
            \end{bmatrix},
\end{equation*}
\begin{equation*}
   D_1 = \begin{bmatrix}
            -0{,}2 & 0{,}8 & -0{,}2
         \end{bmatrix},
    \quad 
    E_1 = \begin{bmatrix}
            0
            \end{bmatrix},
    \quad 
    F_1 =  \begin{bmatrix}
           0{,}02
           \end{bmatrix}.
\end{equation*}
Отметим, что объект является неустойчивым. В этом примере эталонная модель не рассматривается, т.е. все матрицы в \eqref{reference} пустые. Условие \eqref{bounds} предполагается выполненным при $G = 1$.

Пусть динамический порядок $a_3$ регулятора \eqref{controller} положен равным 3, а целевая переменная \eqref{goal} задана как
\begin{equation*}
    z(t) = x(t) - x_c(t),
\end{equation*}
что соответствует задаче наблюдения и выбору матрицы
\begin{equation*}
    K = \begin{bmatrix}
    1 & 0 & 0 & -1 & \phantom{-}0 & \phantom{-}0\\
    0 & 1 & 0 & \phantom{-}0 & -1 & \phantom{-}0\\
    0 & 0 & 1 & \phantom{-}0 & \phantom{-}0 & -1
    \end{bmatrix}.
\end{equation*}
Нетрудно проверить, что в этом случае условие \eqref{x_condition} выполнено. Тогда в результате выполнения алгоритма 1 при $\alpha = 0{,}3$ может быть найдена матрица
\begin{equation*}
    P = \begin{bmatrix}
    \phantom{-}192 &  -624 & \phantom{-}222\\
    -624 & \phantom{..}2051 & -732 \\
    \phantom{-}222 &  -732 & \phantom{-}289
    \end{bmatrix} \succ 0
\end{equation*}
и затем по формулам \eqref{Ylmi}, \eqref{x_solution} восстановлены параметры 
\begin{equation*}
    A_3 = \begin{bmatrix}
    3{,}970 & -15{,}341  &  3{,}750 \\
1{,}506 &  -5{,}585  &  1{,}386   \\
0{,}559  & -1{,}618  &  0{,}349
    \end{bmatrix}, \quad B_3 =
\begin{bmatrix}
19{,}011   \\
6{,}792   \\
1{,}777   
\end{bmatrix}
\end{equation*}
регулятора \eqref{controller}, который в этой задаче выполняет роль наблюдателя. 

Можно заметить, что для полученных матриц выполнено $A_3 = A_1 - B_3D_1$, а значит, структура системы \eqref{controller} совпадает со структурой наблюдателя Люенбергера
\begin{equation*}
    \dot{x_c}(t) = A_1 x_c(t) + B_3(y(t)-D_1x_c(t)).
\end{equation*}
Примечательно, что такая структура не была создана искусственно, а получилась сама собой в результате выполнения алгоритма 1.

При моделировании внешняя помеха была задана как $w(t) = \frac{1}{2}+\frac{1}{2}\operatorname{sgn}(\sin(0{,}1t))$, а начальные условия положены равными $x(0) = \begin{bmatrix} 3{,}2 & 3 & 3 \end{bmatrix}^{\rm T}$ и $x_c(0) = \begin{bmatrix} -10 & 0 & 4\end{bmatrix}^{\rm T}$.

\begin{figure}[H]
\minipage{0.5\textwidth}
  \centering
  \includegraphics[width = \textwidth]{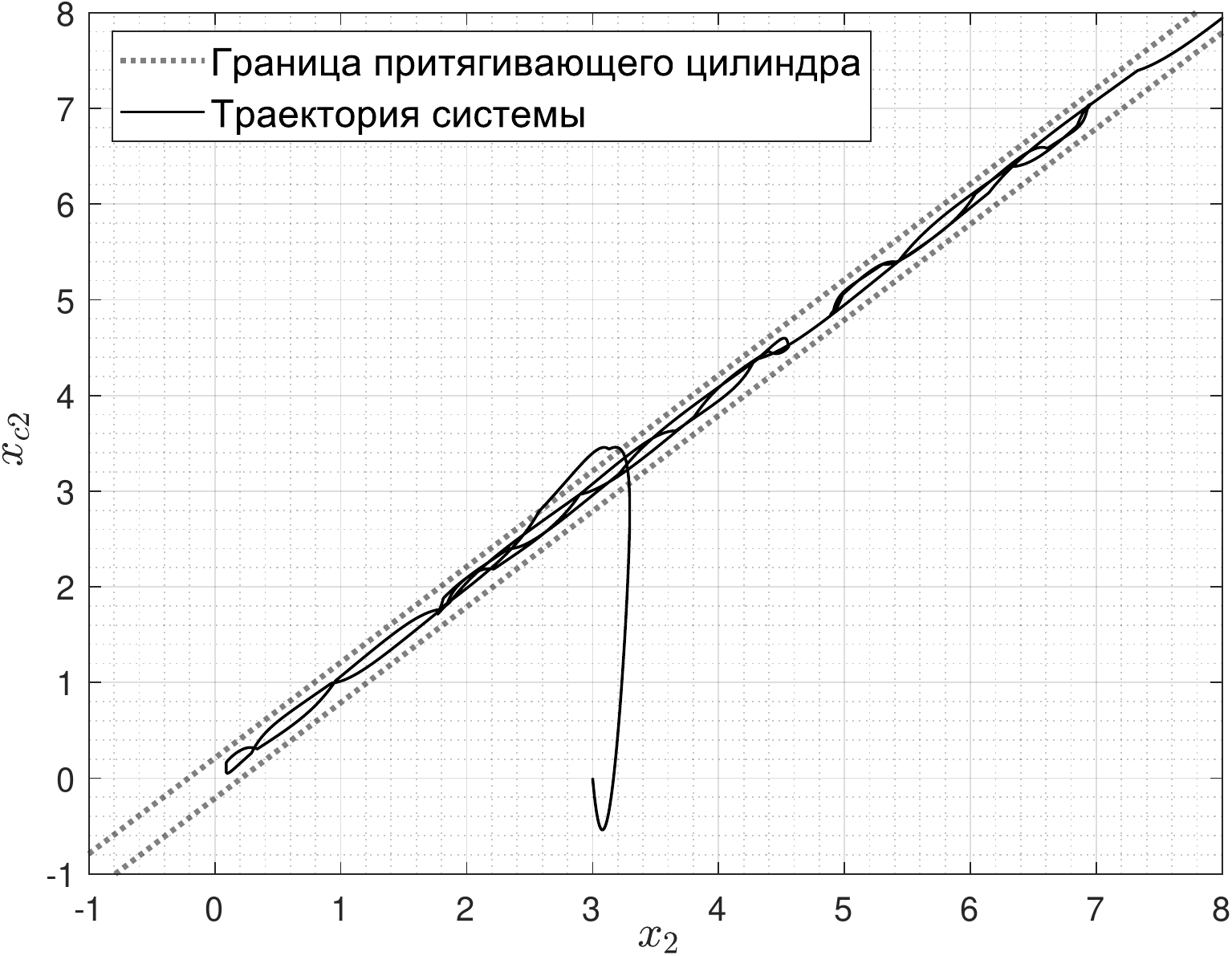}
  \caption{Проекция притягивающего $(k,n)$-цилиндра и траектории замкнутой системы.}
\endminipage\hfill
\minipage{0.5\textwidth}
  \centering
  \includegraphics[width = 0.93\textwidth]{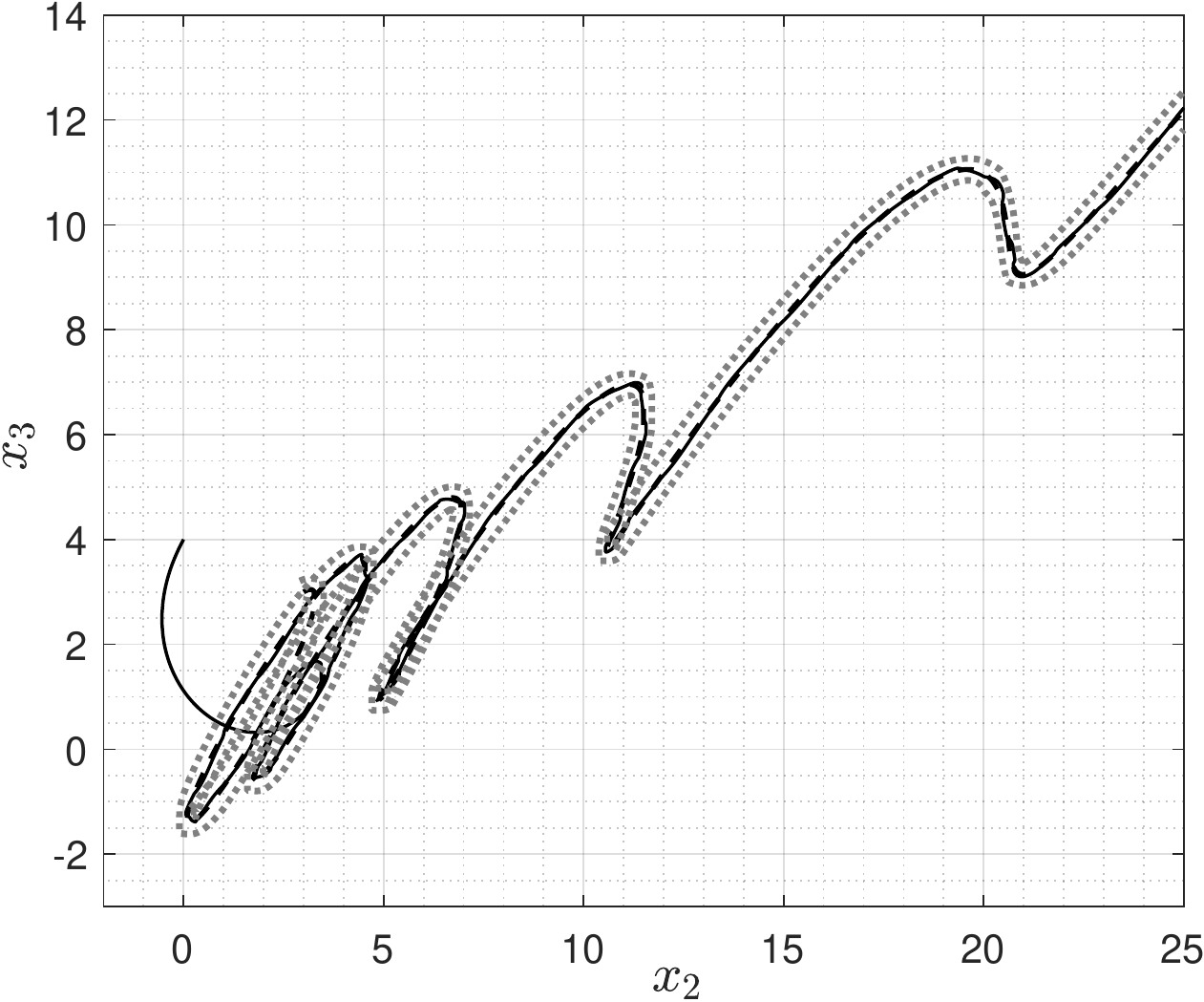}
  \caption{Траектория объекта (\rule[.5ex]{.5em}{.4pt} \rule[.5ex]{.5em}{.4pt} \rule[.5ex]{.5em}{.4pt}), траектория наблюдателя  (\rule[.5ex]{2em}{.4pt})  и 
  границы допустимого коридора (\makebox[2em]{\protect\xdotfill{.4pt}}\!).}
\endminipage
\end{figure}

Результаты моделирования представлены на рис. 7-8. На рис. 7 показана проекция траектории замкнутой системы: по оси абсцисс отложено значение второй координаты вектора состояния объекта, а по оси ординат – значение второй координаты вектора состояния регулятора (наблюдателя). Также показаны границы полосы, являющейся проекцией притягивающего $(k,n)$-цилиндра на эту плоскость.

На рис. 8 показаны траектория $x(t)$ объекта и траектория $x_c(t)$ регулятора (наблюдателя). На рисунке также отмечены границы <<допустимого коридора>>, имеющего следующий смысл: если в данный момент времени каждая из проекций траектории замкнутой системы – на плоскость $(x_2,x_{c2})$ (см. рис. 7) и на плоскость $(x_3,x_{c3})$ (не приводится, так как выглядит аналогично) – находится внутри соответствующей проекции притягивающего $(k,n)$-цилиндра, то траектория $x(t)$ в данный момент времени находится в границах допустимого коридора. Иными словами, тот факт, что траектория наблюдателя начиная с какого-то момента не выходит за границы точечного пунктира, является иллюстрацией того, что траектория замкнутой системы попадает внутрь $(k,n)$-цилиндра и не покидает его.

\section{Заключение}
В работе предложено обобщение метода инвариантных эллипсоидов, позволяющее находить притягивающие подмножества пространства состояний более общего вида. Показано, что предложенный метод может быть использован для решения задач стабилизации, слежения и наблюдения, а также их комбинаций. Предложен алгоритм, позволяющий применять основой результат на практике с помощью стандартных программных средств. На численных примерах продемонстрирована эффективность предложенного подхода. 

Автор выражает благодарность Игорю Борисовичу Фуртату и анонимному рецензенту за ценные замечания, позволившие улучшить форму и содержание статьи.

\appendix{1}  

\begin{proofofstatement}{\ref{st1}}
Матрица $Q$ симметричная, поэтому пространство $\mathbb{R}^n$ можно представить как прямую сумму ее образа и ядра:
\begin{equation*}
    \mathbb{R}^n = \operatorname{range} Q \oplus \operatorname{ker} Q.
\end{equation*}
Иными словами, для любого $x\in \mathbb{R}^n$ существует единственное разложение
\begin{equation*}
    x = x_r + x_k, \quad x_r \in \operatorname{range} Q, \quad x_k \in \operatorname{ker} Q.
\end{equation*}
Тогда $(k,n)$-цилиндр \eqref{opr1} можно представить как 
\begin{equation*}
    \{ x \in \mathbb{R}^n \; | \; x^{\rm T}Qx \le 1\} =   \{ (x_r+x_k) \in \mathbb{R}^n \; | \; (x_r+x_k)^{\rm T}Q(x_r+x_k) \le 1\} =
\end{equation*}
\begin{equation*}
    = \{ (x_r+x_k) \in \mathbb{R}^n \; | \; x_r^{\rm T}Qx_r \le 1\} =  \{ x_r \in \operatorname{range} Q \; | \; x_r^{\rm T}Qx_r \le 1\} + \operatorname{ker} Q.
\end{equation*}
Сужение оператора $Q$ на подпространство $\operatorname{range}Q$ имеет полный ранг. Следовательно, множество $\{ x_r \in \operatorname{range} Q \; | \; x_r^{\rm T}Qx_r \le 1\}$ является $k$-мерным эллипсоидом.
\end{proofofstatement}

\begin{proofofcorollary}{\ref{cor1}}
Символом $\cong$ обозначим гомеоморфность. Из того что $\operatorname{range}Q \cong \mathbb{R}^k$, $\operatorname{ker}Q \cong \mathbb{R}^{n-k}$, а множества $\{ x_r \in \operatorname{range} Q \; | \; x_r^{\rm T}Qx_r \le 1\}$ и $ \operatorname{ker} Q$ ортогональны друг другу, следует, что
\begin{equation*}
     \{ x_r \in \operatorname{range} Q \; | \; x_r^{\rm T}Qx_r \le 1\} + \operatorname{ker} Q \quad \cong \quad \{ x \in \mathbb{R}^k \; | \; x^{\rm T}x \le 1\} \times \mathbb{R}^{n-k}. 
\end{equation*}
\end{proofofcorollary}

\begin{lemma} {\rm \cite{Woodbury}} Если матрицы $A, U, C, V$ таковы, что обе части равенства
\begin{equation*}
(A+UCV)^{-1} = A^{-1}-A^{-1}U(C^{-1}+VA^{-1}U)^{-1}VA^{-1}    
\end{equation*}
имеют смысл (все операции определены), то указанное равенство выполнено. 
\end{lemma}

\begin{lemma} {\rm \cite{Tikhonov}} Для произвольной матрицы $A \in \mathbb{R}^{m \times n}$ верно, что
\begin{equation*}
\lim_{\varepsilon  \,  \to \,  0+}(A^{\rm T}A + \varepsilon I)^{-1}A^{\rm T} = A^+.
\vspace{-0.15cm}
\end{equation*}
Указанное равенство выполнено даже в тех случаях, когда $(A^{\rm T}A)^{-1}$ не существует. 
\end{lemma}

\begin{proofofstatement}{\ref{pq}}
 Воспользуемся полнотой строчного ранга матрицы $C$ и рассмотрим семейство матриц $R_{\varepsilon}$ с параметром $\varepsilon$, определенных как
\begin{equation*}
    R_{\varepsilon} = \Big(C(Q+\varepsilon I)^{-1}C^{\rm T}\Big)^{-1}, \quad \varepsilon > 0.
\end{equation*}
Известно (см. \cite{inv1}), что если $Q \succ 0$, то $R = (CQ^{-1}C^{\rm T})^{-1}$. При $Q \succeq 0$ в силу непрерывности отображения $x \mapsto Cx$ имеем
$R = \lim_{\varepsilon \, \to \, 0+} R_{\varepsilon}$.
Cогласно лемме 1
\begin{equation*}
(Q+\varepsilon I)^{-1} = (\varepsilon I + MM)^{-1} = \frac{1}{\varepsilon}I-\frac{1}{\varepsilon}M(I+\frac{1}{\varepsilon}MM)^{-1}\frac{1}{\varepsilon}M.
\end{equation*}
Тогда
\begin{equation*}
R_{\varepsilon} \, = \Big(C(Q+\varepsilon I)^{-1}C^{\rm T}\Big)^{-1} = \Big( \frac{1}{\varepsilon}CC^{\rm T}-\frac{1}{\varepsilon}CM(I+\frac{1}{\varepsilon}MM)^{-1}\frac{1}{\varepsilon}MC^{\rm T}\Big)^{-1}.
\end{equation*}
Заметим, что матрица $CC^{\rm T}$ обратима. Вновь применяя лемму 1, получаем
\begin{equation*}
R_{\varepsilon} = \varepsilon (CC^{\rm T})^{-1} + (CC^{\rm T})^{-1} CM \Big( I+\frac{1}{\varepsilon}M \big(I - C^{\rm T}(CC^{\rm T})^{-1}C \big) M \Big)^{-1} MC^{\rm T} (C C^{\rm T})^{-1}.
\end{equation*}
Воспользуемся известными равенствами
\begin{equation*}
    C^{\rm T} (C C^{\rm T})^{-1} = C^+, \quad   I - C^{\rm T}(CC^{\rm T})^{-1}C = N = N^2
\end{equation*}
и перепишем $R_{\varepsilon}$ в виде 
\begin{equation*}
    R_{\varepsilon} =\varepsilon (CC^{\rm T})^{-1} + C^{+ \rm T} M \Big( I + \frac{1}{\varepsilon}MNNM\Big)^{-1}MC^+.
\end{equation*}
Применяя лемму 1, получаем 
\begin{equation*}
    R_{\varepsilon} =\varepsilon (CC^{\rm T})^{-1} + C^{+ \rm T} M \Big(   
I-MN \big( \varepsilon I + NMMN \big)^{-1} NM
\Big)MC^+.
\end{equation*}
Тогда согласно лемме 2, учитывая, что $M=M^{\rm T}$, $N = N^{\rm T}$, имеем
\begin{equation*}
    R = \lim_{\varepsilon \, \to \, 0+} R_{\varepsilon} =  C^{+ \rm T} M \Big(I-MN \big( MN \big)^+\Big)MC^+.
\end{equation*}
Наконец заметим, что $I-MN \big( MN \big)^+ \succeq 0$ и, следовательно, $R \succeq 0$.
\end{proofofstatement}
\begin{proofofcorollary}{\ref{pq_plane}}
Если $C$ -- проекция на $\mathbb{R}^2$, то $m = 2$, $r \le 2$. Значит, образом $(k,n)$-цилиндра может быть только $(0,2)$, $(1,2)$ или $(2,2)$-цилиндр.
\end{proofofcorollary}

\begin{lemma} {\rm \cite{lemma1}}
Если $A \in \mathbb{R}^{n \times m}$, $B \in \mathbb{R}^{k \times l}$, $C \in \mathbb{R}^{n \times l}$, то матричное уравнение
\begin{equation*}
AXB = C
\end{equation*}
разрешимо относительно $X \in \mathbb{R}^{m \times k}$ тогда и только тогда, когда
\begin{equation*}
A A^+ C B^+ B = C,
\end{equation*}
и в этом случае все решения можно параметризовать как
\begin{equation*}
X = A^+ C B^+ + Y - A^+ A Y B B^+, 
\end{equation*}
где $Y \in \mathbb{R}^{m \times k}$ -- произвольная матрица.
\end{lemma}

\begin{lemma} {\rm \cite{lemma2}}
Если $A \in \mathbb{S}^n$, $B \in \mathbb{R}^{m \times n}$ и $A^2 = A$, то $A(BA)^+ = (BA)^+$.
\end{lemma}

\begin{lemma} {\rm \cite{lemma3}}
Если $A \in \mathbb{R}^{n \times m}$, $B \in \mathbb{R}^{k \times n}$, $C \in \mathbb{S}^n$, то матричное неравенство 
\begin{equation*}
    AXB+(AXB)^{\rm T}+C \prec 0
\end{equation*}
разрешимо относительно $X \in \mathbb{R}^{m \times k}$ в том и только в том случае, если существуют $\mu_1, \mu_2 \in \mathbb{R}$ такие, что
\begin{equation*}
    C \prec \mu_1 A A^{\rm T}, \quad C \prec \mu_2 B^{\rm T} B.
\end{equation*}

\end{lemma}

\begin{proofoftheorem}{\ref{thm:1}}
Введем переменную $y(t) = Cx(t) \in \mathbb{R}^k$, составим уравнение ее динамики \color{black}
\begin{equation*}
\dot{y}(t) = CAx(t)+CBf(t)  
\end{equation*}
и найдем условие, при котором оно может быть записано независимо от $x(t)$, т.е. условие существования матрицы $X$ такой, что
\begin{equation*}
     \dot{y}(t)  = XC x(t) + CBf(t)= X y(t) + CBf(t).  
\end{equation*} 
\color{black}Для этого рассмотрим уравнение $CA = XC$. В соответствии с леммой 3 оно разрешимо относительно $X$ в том и только в том случае, если выполнено условие $CA(I-C^+C)=0$, и тогда все решения могут быть параметризованы как $X = CAC^+ + Y(I-CC^+)$, где $Y$ -- произвольная матрица соответствующей размерности. Если дополнительно выполнено условие $\operatorname{rank} C = k$, то уравнение имеет единственное решение $X = CAC^+$.

Так как все соответствующие условия включены в формулировку теоремы, динамика переменной $y(t)$ может быть записана независимо от $x(t)$ в виде
\begin{equation} \label{proof_y}
\dot{y}(t) = CAC^+ y(t)+CBf(t).    
\end{equation}
Обозначим: $V = y^{\rm T}Py$. Заметим, что из матричного неравенства
\begin{equation*}
\begin{bmatrix}
PCAC^{+}+(CAC^{+})^{\rm T}P+\alpha P & PCB \\ (CB)^{\rm T}P & -\alpha G 
\end{bmatrix} \prec 0
\end{equation*}
следует, что 
\begin{equation*}
\begin{bmatrix}
y(t) \\ f(t)
\end{bmatrix} ^{\rm T}
\begin{bmatrix}
PCAC^{+}+(CAC^{+})^{\rm T}P+\alpha P & PCB \\ (CB)^{\rm T}P & -\alpha G 
\end{bmatrix} \begin{bmatrix}
y(t) \\ f(t)
\end{bmatrix} \le 0, \quad \forall t \ge 0,
\end{equation*}
и тогда
\begin{equation*}
\dot{V}(t) + \alpha V(t) \le \alpha f^{\rm T}(t)Gf(t), \quad \forall t \ge 0.
\end{equation*}
При выполнении условия \eqref{anal2} из этого следует, что 
\vspace{-0.1cm}
\begin{equation*}
\vspace{-0.1cm}
\limsup_{t \rightarrow +\infty} V(t) \le 1
\end{equation*}
и если $V(t_0) \le 1$, то $V(t) \le 1$ при всех $t \ge t_0$. При этом $V =  x^{\rm T}C^{\rm T}PCx$, $C^{\rm T}PC \succeq 0$ и $\operatorname{rank} C^{\rm T}PC = k$, а значит, подмножество 
\vspace{-0.1cm}
\begin{equation*}
\Big\{x \in \mathbb{R}^n  \; \Big| \;  x^{\rm T}Qx \le 1 \Big\}, \quad Q = C^{\rm T}PC
\vspace{-0.1cm}
\end{equation*}
пространства состояний системы \eqref{anal1}-\eqref{anal2} является притягивающим $(k,n)$-цилиндром.
\end{proofoftheorem}
\vspace{-0.5cm}
\begin{proofoftheorem}{\ref{thm:2}}
В соответствии с теоремой \ref{thm:1} подмножество $\{s \in \mathbb{R}^{n}  \; | \;  s^{\rm T}K^{\rm T}PKs \le 1\}$ системы \eqref{closed} является притягивающим $(k,n)$-цилиндром, если выполнены условия
\begin{equation} \label{cond1}
    KM(I-K^+K)=0, 
\end{equation}
\begin{equation} \label{cond2}
\begin{bmatrix}
PKMK^{+}+(KMK^{+})^{\rm T}P+\alpha P & PKN \\ (KN)^{\rm T}P & -\alpha G 
\end{bmatrix} \prec 0.
\end{equation}

Для существования регулятора \eqref{controller} такого, чтобы для замкнутой системы \eqref{closed}  выполнялось условие \eqref{cond1}, необходимо и достаточно, чтобы уравнение
\begin{equation*}
    K(A+BXD)(I-K^+K) = 0 \quad \Leftrightarrow \quad  KBXD(I-K^+K) = -KA(I-K^+K)
\end{equation*}
было разрешимо относительно $X$. В соответствии с леммой 3 это так в том и только в том случае, если выполнено
\begin{equation} \label{x_condition_proof}
    KB(KB)^+ KA(I-K^+K) ( D(I-K^+K))^+  D(I-K^+K) = KA(I-K^+K),
\end{equation}
причем все соответствующие матрицы $X$ могут быть параметризованы как
\begin{equation}\label{x_solution_proof}
\begin{aligned}
    X = &(KB)^+ KA(K^+K-I) (D(I-K^+K))^+ \\ & \quad \quad + Y - (KB)^+KBYD(I-K^+K)(D(I-K^+K))^+,
\end{aligned}
\end{equation}
где $Y$ -- произвольная матрица соответствующей размерности.
Согласно лемме 4 \eqref{x_condition_proof} и \eqref{x_solution_proof} равносильны \eqref{x_condition} и \eqref{x_solution} соответственно.

С учетом выражений $M = A+BXD$, $N = C + BXF$ условие \eqref{cond2} может быть переписано в виде
\begin{equation*}
    \begin{bmatrix}
    PKAK^+ + (KAK^+)^{\rm T}P + \alpha P & PKC \\ (KC)^{\rm T}P & -\alpha G  \end{bmatrix} + \hspace{5cm}
    \end{equation*}
    \begin{equation*}
    \hspace{3cm} +
    \begin{bmatrix}
    PKB \\ 0  \end{bmatrix} X \begin{bmatrix}  DK^+ & F  \end{bmatrix} + \Big(\begin{bmatrix}
    PKB \\ 0  \end{bmatrix} X \begin{bmatrix}  DK^+ & F  \end{bmatrix}\Big)^{\rm T} \prec 0
\end{equation*} 
и после подстановки \eqref{x_solution} и применения обозначений \eqref{notations} представлено как
\begin{equation} \label{Ylmi_proof}
    \begin{bmatrix}
    P H_1  + H_1^{\rm T} P + \alpha P & P H_2 \\ H_2^{\rm T} P  & - \alpha G
    \end{bmatrix} + \begin{bmatrix}
    PH_3 \\ 0
    \end{bmatrix} Y \begin{bmatrix}
    H_4 & H_5
    \end{bmatrix} + \Big( \begin{bmatrix}
    PH_3 \\ 0
    \end{bmatrix} Y \begin{bmatrix}
    H_4 & H_5
    \end{bmatrix} \Big)^{\rm T}
    \prec 0.
\end{equation}
Согласно лемме 5 соответствующая матрица $Y$ -- а значит, и набор $X$ параметров регулятора \eqref{controller} -- существует в том и только в том случае, если найдутся $\mu_1, \mu_2 \in \mathbb{R}$, при которых выполнены матричные неравенства
\begin{equation} \label{lmi1_not}
    \begin{bmatrix}
    P H_1  + H_1^{\rm T} P + \alpha P & P H_2 \\ H_2^{\rm T} P  & - \alpha G
    \end{bmatrix} \prec \mu_1 \begin{bmatrix}
    P H_3 H_3^{\rm T} P & 0 \\ 0 & 0 
    \end{bmatrix}, 
\end{equation}
\vspace{-0.5cm}
\begin{equation} \label{lmi2}
    \; \; \begin{bmatrix}
    P H_1  + H_1^{\rm T} P + \alpha P & P H_2 \\ H_2^{\rm T} P  & - \alpha G
    \end{bmatrix} \prec \mu_2 \begin{bmatrix}
    H_4^{\rm T} H_4 & H_4^{\rm T} H_5 \\ H_5^{\rm T} H_4 & H_5^{\rm T} H_5
    \end{bmatrix}.
\end{equation}
Заметим, что если матрица $Q$ такова, что $PQ=I$, то \eqref{lmi1_not} равносильно 
\begin{equation} \label{lmi1}
    \hspace{-0.6cm}\begin{bmatrix}
    H_1 Q + Q H_1^{\rm T} + \alpha Q & H_2 \\ H_2^{\rm T} & - \alpha G
    \end{bmatrix} \prec \mu_1 \begin{bmatrix}
    H_3 H_3^{\rm T} & 0 \\ 0 & 0 
    \end{bmatrix},
    \vspace{-0.3cm}
\end{equation}
чтобы увидеть это, достаточно умножить \eqref{lmi1_not} слева и справа на $\begin{bmatrix}
Q & 0 \\ 0 & I
\end{bmatrix} \succ 0$.

Для завершения доказательства осталось заметить, что \eqref{lmi2}, \eqref{lmi1} совпадают с \eqref{PQlmi}, \eqref{Ylmi_proof} равносильно \eqref{Ylmi} и что при фиксированных $\alpha, P$ матричное неравенство \eqref{Ylmi} является линейным.
\end{proofoftheorem}

\end{document}